\documentclass[10pt,hidelinks,a4paper,twocolumn]{article}
\usepackage{xcolor}
\usepackage{subcaption}
\usepackage{verbatim}
\usepackage{siunitx}
\usepackage{graphicx}
\usepackage{authblk}
\usepackage{hyperref}
\usepackage{abstract}

\newcommand*{\wl}[1]{\lambda_{\mathrm{#1}}}
\newcommand*\chem[1]{\ensuremath{\mathrm{#1}}}

\title{High-birefringence direct UV-written waveguides for heralded single-photon sources at telecommunication wavelengths}

\author[1,*]{Matthew T.\ Posner}
\author[2]{T. Hiemstra}
\author[1]{Paolo L. Mennea}
\author[1]{Rex H.\ S.\ Bannerman}
\author[2]{Ulrich B.\ Hoff}
\author[2]{Andreas Eckstein}
\author[2,3]{W.\ Steven Kolthammer}
\author[2]{Ian A. Walmsley}
\author[1]{Devin H.\ Smith}
\author[1]{James C. Gates}
\author[1]{Peter G.\ R.\ Smith}
\affil[1]{Optoelectronics Research Centre, University of Southampton, University Rd, Southampton, SO17 1BJ, UK}
\affil[2]{Clarendon Laboratory, University of Oxford,  Parks Rd, Oxford, OX1 3PU, UK}
\affil[3]{Blackett Laboratory, Imperial College London, Prince Consort Rd, SW7 2AZ, UK}
\affil[*]{\href{mailto:m.posner@soton.ac.uk}{\nolinkurl{m.posner@soton.ac.uk}}}
\date{\vspace{-6ex}}
 
\begin{document}
\twocolumn[

\maketitle
\begin{onecolabstract}
Direct UV-written waveguides are fabricated in silica-on-silicon with birefringence of \num{4.9+- 0.2e-4}, much greater than previously reported in this platform. We show that these waveguides are suitable for the generation of heralded single photons at telecommunication wavelengths by spontaneous four-wave mixing. A pulsed pump field at \SI{1060}{nm} generates pairs of photons in highly detuned, spectrally uncorrelated modes near \SI{1550}{nm} and \SI{800 }{nm}. Waveguide-to-fiber coupling efficiencies of \SIrange{78}{91}{\percent} are achieved for all fields. Waveguide birefringence is controlled through dopant concentration of \chem{GeCl_4} and \chem{BCl_3} using the flame hydrolysis deposition process. The technology provides a route towards the scalability of silica-on-silicon integrated components for photonic quantum experiments.
\end{onecolabstract}]

\section{Introduction}

Applications in photonic quantum information and quantum communications require low loss: silica photonics provides a promising 
platform for these applications \cite{Orieux2016}. As such, silica-based photonic integrated circuits offer a route to scalable and reproducible technology for integration of different components in a complex quantum network\cite{Politi2011}. 
Sources of heralded single photons have been achieved in waveguides on a silica chip through birefringent phase-matched spontaneous four-wave mixing (SFWM). Highly uniform waveguides have enabled an array of nearly identical sources, with low loss and high spectral purity \cite{Spring2017}.
Phase matching in highly birefringent waveguides can generate pairs of photons with one in the near-infrared (near-IR) (\SIrange{800}{830}{nm}) and the other in the telecommunications C-band (\SIrange{1530}{1570}{nm}) which can thus  be readily separated and isolated from the pump light. Furthermore, the large spectral separation minimizes noise from spontaneous Raman scattering\cite{Lin2007}, 
contributing to improved heralding efficiency and generation rates.

\begin{figure*}[bt]
\centering
\includegraphics[width=0.8\textwidth]{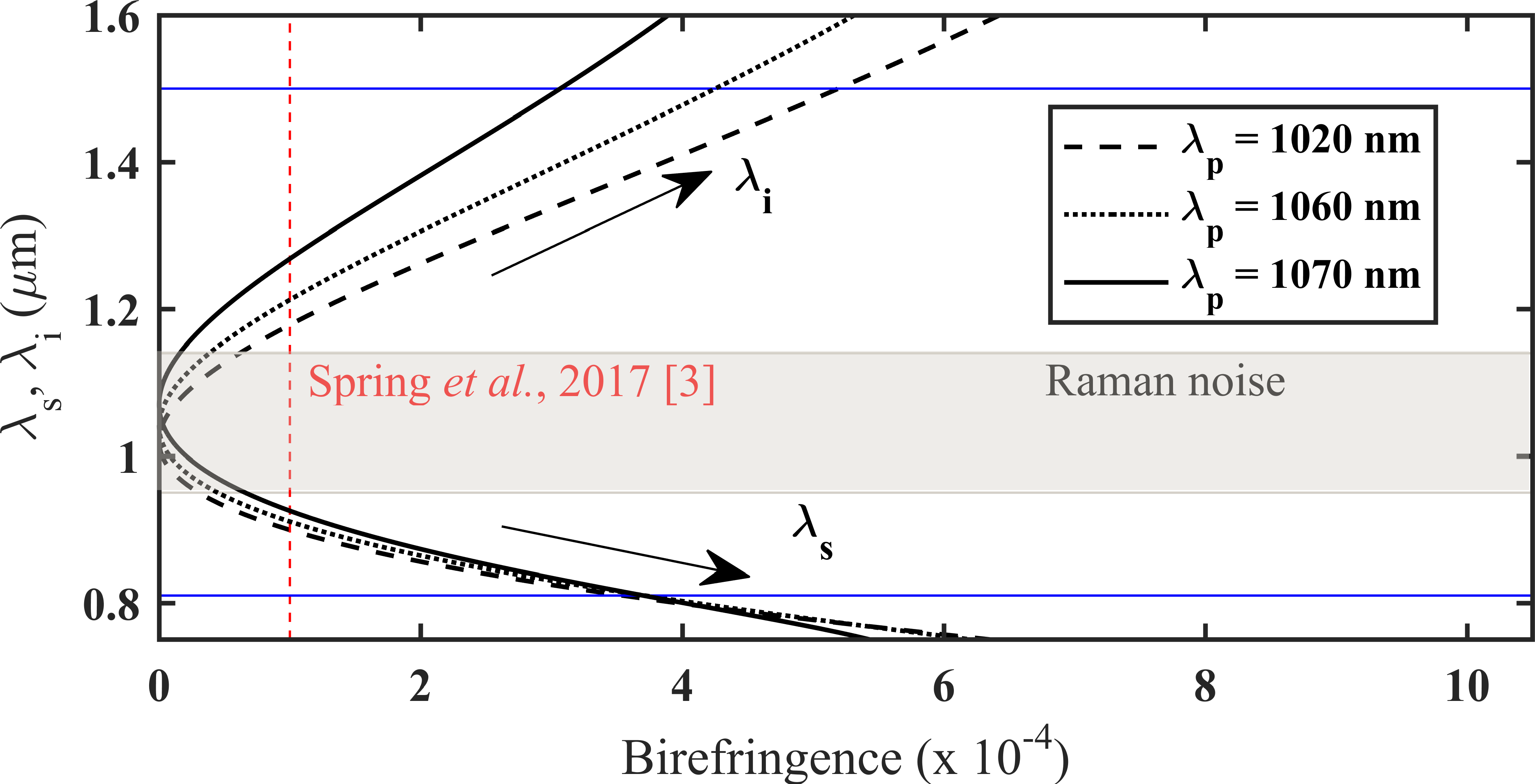}
\caption{Phase-matched solutions of Eq.~(\ref{eq:phasematching})
as a function of birefringence at pump wavelengths $\wl{p}$ of \SIlist{1020;1060;1070}{nm}.  The lower and upper branches of the curves represent the signal and idler wavelength $\wl{s}$ and $\wl{i}$, respectively. The shaded area indicates the range of Raman noise resulting from the pump wavelengths considered. In blue are target optimal wavelengths for low-loss fiber compatible idler photons and near-IR heralding photons. 
}
\label{fig:phasematching}
\end{figure*}

Lithographic processes \cite{Politi2008} and femtosecond-laser (\si{fs}-laser) writing \cite{Marshall2009} are techniques that have been demonstrated for the fabrication of waveguide-based silica photonic quantum circuits. For \si{fs}-laser written waveguides, modal shaping can be employed to generate waveguides with high birefringence. In previous investigations of this approach, however, lack of uniformity of the birefringence in waveguide arrays has limited the scalability\cite{Spring2013} and, to the best of our knowledge, there are no reports of arrays of silica SFWM sources in \si{fs}-laser written waveguides with high birefringence. It is well known that etching used for lithographic processes results in waveguides with low birefringence due to the release of layer stress \cite{Kawachi1990, Dai2005}, which is incompatible with achieving the high-birefringence waveguides required for large spectral detuning and to overcome Raman noise in the photon generation process.

Direct UV-writing (DUW)\cite{Sima2013} is an alternative laser-based fabrication technique that has been used to demonstrate arrays of near-identical waveguide-sources using SFWM for the generation of heralded single-photon pairs centered at wavelengths of 670 and \SI{817}{nm}\cite{Spring2017}. The process permits the fabrication of buried channel waveguides in a photosensitive slab silica layer with high uniformity of waveguide birefringence and mode profile, low insertion loss to silica-fiber networks, and the possibility of integrated classical diagnostics. All of these features make DUW an attractive technique for further scaling and integration of these sources. To date however, high-birefringence waveguides have not been demonstrated using this platform. 

This study investigates the use of flame hydrolysis deposition (FHD) 
to create high-birefringence DUW waveguides in a silica-on-silicon (SoS) platform, whereas prior work in this platform has largely concentrated on methods to reduce and eliminate the birefringence\cite{Okuno1994,Kilian2000,Holmes2015}. High-birefringence waveguides have been reported by integrating metal dopants or stress applying films \cite{Kawachi1990}; 
both approaches result in undesirable losses \cite{Kawachi1990} and consequently have not been considered here. Birefringence in FHD silica waveguides 
is largely caused by thermal-expansion mismatch between the different materials in a layered slab waveguide. This mismatch causes stress in the layers, which in turn introduces an intrinsic birefringence within the layers\cite{Huang:03}.
In the context of planar silica fabrication, it is well understood that a reduction in the concentration of germanium and boron dopants in silica leads to a decrease in the glass thermal expansion coefficient and an increase in the layer's stress \cite{Sakaguchi1994}. FHD allows for individual control of the dopant concentration in each layer of the device; here we study the effect of controlling the material composition to enable the production of highly birefringent layers.

This paper presents waveguide design and fabrication methods for high-bire\-frin\-gence DUW waveguides in FHD wafers. Characterization methods adapted from DUW in-situ grating interrogation \cite{Rogers2012} and seeded joint-spectral intensity measurements \cite{Eckstein2014} are used to classically determine birefringence and assess the quality of the spectral correlations of the fabricated devices. Additionally, the efficiency of direct waveguide-to-fiber coupling is investigated, as it is an important factor in determining the platform's scalability.

\section{SFWM heralded single-photon sources in silica-on-silicon platforms}

SFWM in silica occurs due to a third order non-linear process\cite{Agrawal2013397}. Two photons from a pump pulse are converted to a pair of daughter photons, known as signal and idler, under the conditions of phase-matching and energy conservation.
Birefringence in waveguides modifies the phase-matching condition, increasing the spectral separation between signal and idler\cite{Agrawal2013397}.  
Pair generation is non-deterministic, and while the small third-order nonlinear susceptibility of silica limits the generation rate \cite{Eisaman2011}, mode confinement in a waveguide provides a counteracting enhancement \cite{Spring2017}.
Moreover, the generated photon pairs are contained in well-defined waveguide modes that can be transferred to an optical fiber with minimal coupling loss due to the deliberately designed close match in their mode profile. 

The spectral position of the signal and idler photons can be predicted as a function of the pump wavelength using the wave-vector mismatch in birefringent waveguides, taking into account the cross-polarized SFWM scheme employed in Smith \textit{et al.}\cite{Smith2009}. 
The wave-vector mismatch, $\Delta k$, equal to zero in the phase-matched case, is given by
\begin{equation}
\label{eq:phasematching}
 \frac{\Delta k}{2\pi} = 2\frac{n(\wl{p})}{\wl{p} }
 -  \frac{n(\wl{s})}{\wl{s}} - \frac{n(\wl{i})}{\wl{i}} + 2 \frac{\Delta n}{\wl{p}},
\end{equation}
where $\Delta n$ is the birefringence of the waveguide, treated as a constant perturbation of the material properties.  The subscripts $\{\mathrm{p,s,i}\}$ refer to the pump, signal and idler fields respectively, with $\lambda$ the vacuum wavelength of each field. The index of refraction,
$n(\lambda)$,  has been determined by multi-order interrogation of DUW-written Bragg gratings, fitted to the Sellmeier equation for our FHD glass \cite{Rogers2012}.

The solutions of Eq.~\ref{eq:phasematching} as a function of $\Delta n$ for the phase-matched condition with energy conservation are shown in Fig. \ref{fig:phasematching} for three pump wavelengths in the range of \SIrange{1020}{1070}{nm}, which conveniently matches the spectral range of compact Ytterbium-fiber laser systems. The shaded region 
depicts the \SI{12 }{THz} range of Raman noise\cite{Stolen1980}, which poses a challenge for SFWM sources.
The DUW waveguides used in Spring \textit{et al.}, 2017, \cite{Spring2017} have a birefringence of \num{1e-4}, indicated by the vertical dashed line in Fig. \ref{fig:phasematching}.
Increasing the 
birefringence further greatly reduces the filtering required to remove the Raman noise from the signal and idler as they move farther from the peak of the noise band. 
For waveguides with birefringence greater than \num{4e-4}, the platform generates photon-pairs that are in the near-IR and telecommunications C-band,  which are desirable wavelengths for silicon photodetectors and low-loss telecommunication-band photonics, respectively. 

\section{Fabrication}

To fabricate planar silica layers with different material compositions, the commercial standard process of FHD was used. The process synthesizes fine glass particles that are deposited as a porous low-density soot on a planar wafer. The formation of the silica soot takes place through the combination of the oxidation and hydrolysis of silicon tetrachloride (\chem{SiCl_4}) in an oxyhydrogen burner. An admixture of other chlorides is used to add dopants during layer production, directly controlling the dopant concentration in the deposited soot. The  delivery of dopants to the deposition chamber is controlled by a mass flow controller and a bubbler control system; 
increasing the gas flow increases the volume of precursor reaching the burner. The dopant concentration tunes the properties of the FHD glass, including the refractive index, thermal expansion coefficient and melting point.

The dopants used in this study to modify the thermal expansion coefficient of the core layer are germanium tetrachloride (\chem{GeCl_4}) and boron trichloride (\chem{BCl_3}). All other processing parameters are identical to those of the devices used in Spring \textit{et al.}, 2017 \cite{Spring2017}. The soot was deposited on base wafers comprised of silicon substrates with a \SI{15}{\micro m} thick thermal oxide layer. This oxide layer acts as a buffer to alleviate the stresses between the silicon and the silica and also as an optical under-cladding for the waveguides. Following soot deposition, the wafers are transferred to a furnace 
for consolidation. A subsequent FHD step to deposit and consolidate soot for an over-cladding layer is carried out after the core layer fabrication.

\begin{figure*}[tbh]
\centering\includegraphics[width=0.90\textwidth]{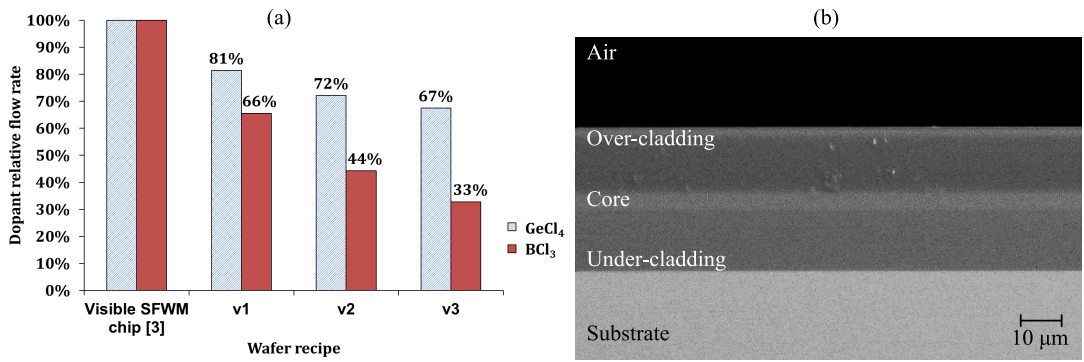}
\caption{(a) \chem{GeCl_4} and \chem{BCl_3} relative flow rates in wafers relative to near-IR heralded single-photon source wafer used in Spring \textit{et al.}, 2017 \cite{Spring2017}. 
(b) Representative backscattered electron microscope image of end facet of a fabricated chip.}
\label{fig:fab}
\end{figure*}

Three SoS wafers with differing compositions of boron and germanium in the core planar layers were fabricated. The FHD dopant flow rates of \chem{GeCl_4} and \chem{BCl_3} relative to those of wafers used in Spring \textit{et al.}, 2017, \cite{Spring2017} 
are shown in Fig.~\ref{fig:fab} (a); these will subsequently be referred to as wafers v1, v2 and v3. Each wafer underwent independent FHD and consolidation steps to initially generate a germano\-boro\-silicate photosensitive core layer followed by a boro\-phospho\-silicate over-cladding layer. The temperatures used for the consolidation of the core and cladding layer were \SI{1360}{\celsius}  and \SI{1100}{\celsius}, respectively. The lower consolidation temperature of the over-cladding layer prevents the re-flowing of the previously consolidated core layer. Additional control wafers for each layer deposited were also fabricated to test the optical and material properties of the layers. 

The FHD layers were characterized by a prism coupling measurement technique (Metricon 2010): the core layers had a refractive index contrast of \SI{0.2}{\percent} with respect to the cladding; furthermore, the over-cladding had a refractive index matched to that of the thermal oxide under-cladding at \SI{1553}{nm}. The chips were diced and the end facets inspected by electron microscopy to assess the layer quality. Figure~\ref{fig:fab} (b) shows distinct glass layers in a fabricated chip, which is representative of layers in all three fabricated wafers. The chips were kept in a hydrogen atmosphere at\SI{120}{ \bar} for at least 240 hours to enhance photosensitivity prior to DUW.

DUW was used to fabricate the waveguides and gratings simultaneously into the photosensitive FHD silica core layer. The irradiation source was a frequency-doubled Argon-ion laser ($\lambda= \SI{244}{nm}$), 
which was split into two coherent beams that were focused and overlapped into a small spot (\SI{6}{\um} diameter) \cite{Sima2013}. Through electro-optic modulation of one arm of this interferometer, the interference pattern resulting from the overlap of the two beams was used to form a Bragg grating while writing the waveguide; in addition, phase modulation control of the interferometer arm was used to detune the Bragg grating's central wavelength\cite{Sima2013}.
The gratings were used to measure the birefringence along the waveguide, as will be discussed in section~\ref{characterization}. 

Waveguide-chips from three different wafers with varying core layer compositions were made by DUW. 
Each chip was comprised of eight waveguides with exposure fluence from \SIrange[range-phrase = { to }]{10}{20}{kJ/cm^2} to study waveguide formation by DUW, as well as three control waveguides to study the birefringence variation as a function of time and waveguide position on the chip. Every waveguide contained six diagnostic gratings, each with different wavelength, 
made by DUW\cite{Sima2013}.

\section{Waveguide characterization} \label{characterization}

\begin{figure*}[tb]
\centering\includegraphics[width=0.95\textwidth]{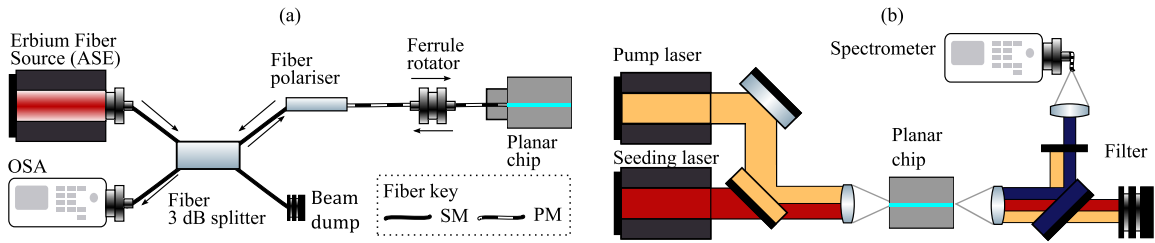}
\caption{Characterization setups for integrated high-birefringent waveguide sources. (a) Fiber-coupled interrogation of TE and TM back-reflected radiation from on-chip DUW waveguide-gratings. (b) Free-space joint spectral intensity measurement for characterization of spectral correlations in the SFWM process. SM: single-mode. PM: polarization-maintaining. 
}
\label{fig:setups}
\end{figure*}

To guide the fabrication study, as well as to predict the quantum behaviour of the devices, characterization of the waveguide properties was performed. Knowledge of the waveguide birefringence is essential to predict the signal and idler wavelengths in birefringent phase-matched SFWM heralded single-photon sources, as shown by Eq.~(\ref{eq:phasematching}). Further, to estimate the capability of photon sources for high interference visibility, joint spectral measurements of the output state can be performed to study its spectral correlations. Waveguide birefringence and spectral correlations have been measured by adapting techniques for DUW in-situ grating interrogation \cite{Rogers2012} and seeded joint-spectral intensity (JSI) measurements \cite{Eckstein2014} using the setups shown in Fig.~\ref{fig:setups} (a) and (b), respectively. Methods and results are described in this section. 

The measurement of the waveguide birefringence was performed by the interrogation of the gratings' spectral reflection properties for both TE and TM polarized waveguide modes. Each grating's reflectivity spectra provides information on the reflected mode's propagation constant, $\beta = 2\pi n_{\textrm{eff}}/\lambda$, for each polarization state at the position of the grating. This technique allows birefringence mapping along a waveguide without resorting to destructive cut-back techniques.  

The reflected spectrum from the gratings in each waveguide for the TE and TM polarization were collected using the setup illustrated in Fig.~\ref{fig:setups} (a). An amplified spontaneous emission (ASE) erbium-fiber source operating in the C-band was used as the input. The polarization of the launch signal (TE, TM) was controlled through a fiber polarizer and a fiber ferrule rotator. The signal was coupled on chip using a fiber mounted in a V-groove assembly and the reflected spectrum from the Bragg gratings was measured with an optical spectrum analyzer (OSA). The unused arm of the fiber beam-splitter was terminated with refractive index matching gel to increase the signal-to-noise ratio of the measurement by reducing the back-reflection. Each waveguide was tested successively by measuring the grating reflectivity for both polarizations. 

\begin{figure*}[tb]
 \begin{subfigure}[t]{0.5\textwidth}
 \caption{}\label{fig:reflectivity}
 \centering
   \includegraphics[width=0.9\textwidth]{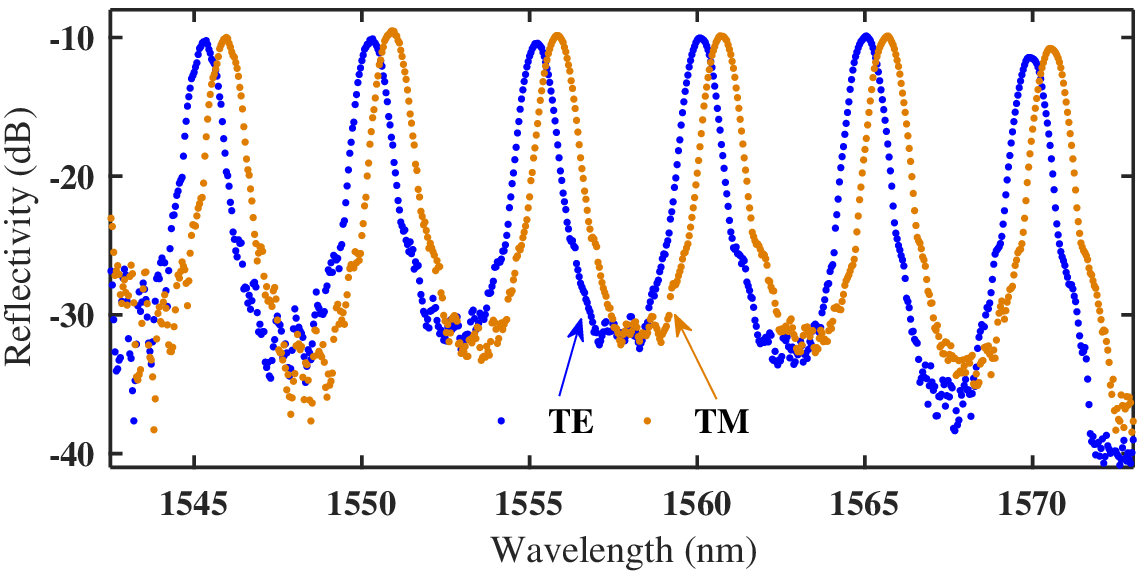}
\end{subfigure}
\begin{subfigure}[t]{0.5\textwidth}
\caption{}\label{fig:birefringence}
  \includegraphics[width=0.9\textwidth]{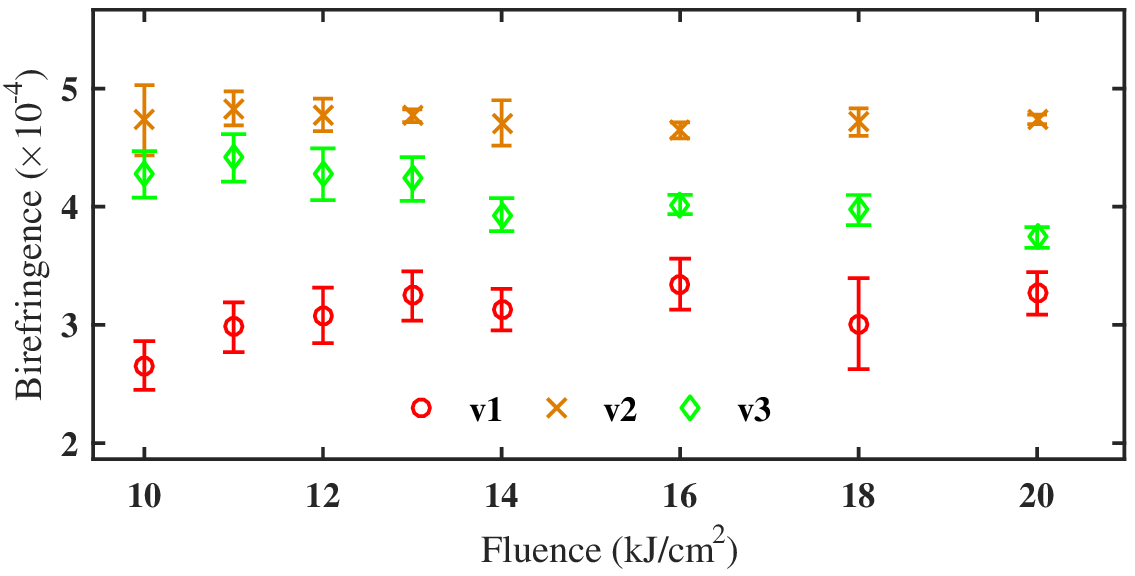}
\end{subfigure}
\caption{Bragg grating characterization results. (\subref{fig:reflectivity}) Representative Bragg grating reflection spectrum for TE and TM modes of high-birefringence DUW waveguides from wafer v2. (\subref{fig:birefringence}) Waveguide birefringence as a function of fluence for waveguides on three chips from wafers v1, v2 and v3.   
}
\label{fig:gratingsResults}
\end{figure*}

Figure \ref{fig:gratingsResults} (\subref{fig:reflectivity}) illustrates typical reflection spectra of TE and TM modes for Bragg gratings in DUW waveguides with high-birefringence. The birefringence at the central position of the grating is directly obtained by computing the cross-correlation of TE and TM reflected signals of each grating. The waveguide birefringence is calculated as the average value of the birefringence measured from all six gratings. Figure~\ref{fig:gratingsResults} (\subref{fig:birefringence}) shows the average waveguide birefringence as a function of DUW fluence for three chips from wafers v1, v2 and v3. The error bars indicate the standard deviation calculated from the six different gratings in each waveguide. The fluence does not significantly change the waveguide birefringence, indicating that the birefringence is dominated by the intrinsic stress of the FHD core layer.  

The birefringence of the waveguides measured on the chips from wafer v2 yield the highest birefringence. This composition is suitable to achieve the large detuning required for the generation of signal and idler fields in the near-IR and C-band, as predicted by Eq.~(\ref{eq:phasematching}). 
The chip's birefringence uniformity was assessed using the control waveguides, with for all three waveguides $\Delta n = \num{4.5+-0.2e-4}$.
The data from wafer v3 suggest that further dopant reduction may not yield a higher birefringence, setting an approximate upper bound for the birefringence that can be achieved using this approach. 

Two additional waveguide-chips from different areas of wafer v2 were subsequently fabricated by DUW, each with source-waveguides and three control waveguides. The source-waveguides did not have gratings, which could interfere with the generated idler field.  
Using the Bragg gratings within the control waveguides, the average birefringence of each chip was measured to be $\Delta n = \num{4.5+-0.2e-4}$ and $\Delta n = \num{4.9+-0.2e-4}$. Variation of position on the \SI{150}{mm} wafer is a likely cause for the variation in birefringence between chips; however, the variation across the chip was small.

The joint spectral distribution of photon pairs created by SFWM was investigated using stimulated four-wave mixing;
the measurement uses a continuous-wave laser scanned over the range of the idler spectrum. This, in combination with the pump beam, leads to stimulated emission in the signal field, amplifying the otherwise weak signal; the amplified signal field can be detected on a spectrometer \cite{Eckstein2014}. The seeded measurement's spectral distribution is identical to the joint spectrum of the device in single-photon operation up to variations in the power of the seed beam.

\begin{figure*}[tbh]
 \begin{subfigure}[c]{0.5\textwidth}
 \caption{}\label{fig:JSI1}
  \centering
   \includegraphics[width=0.85\textwidth]{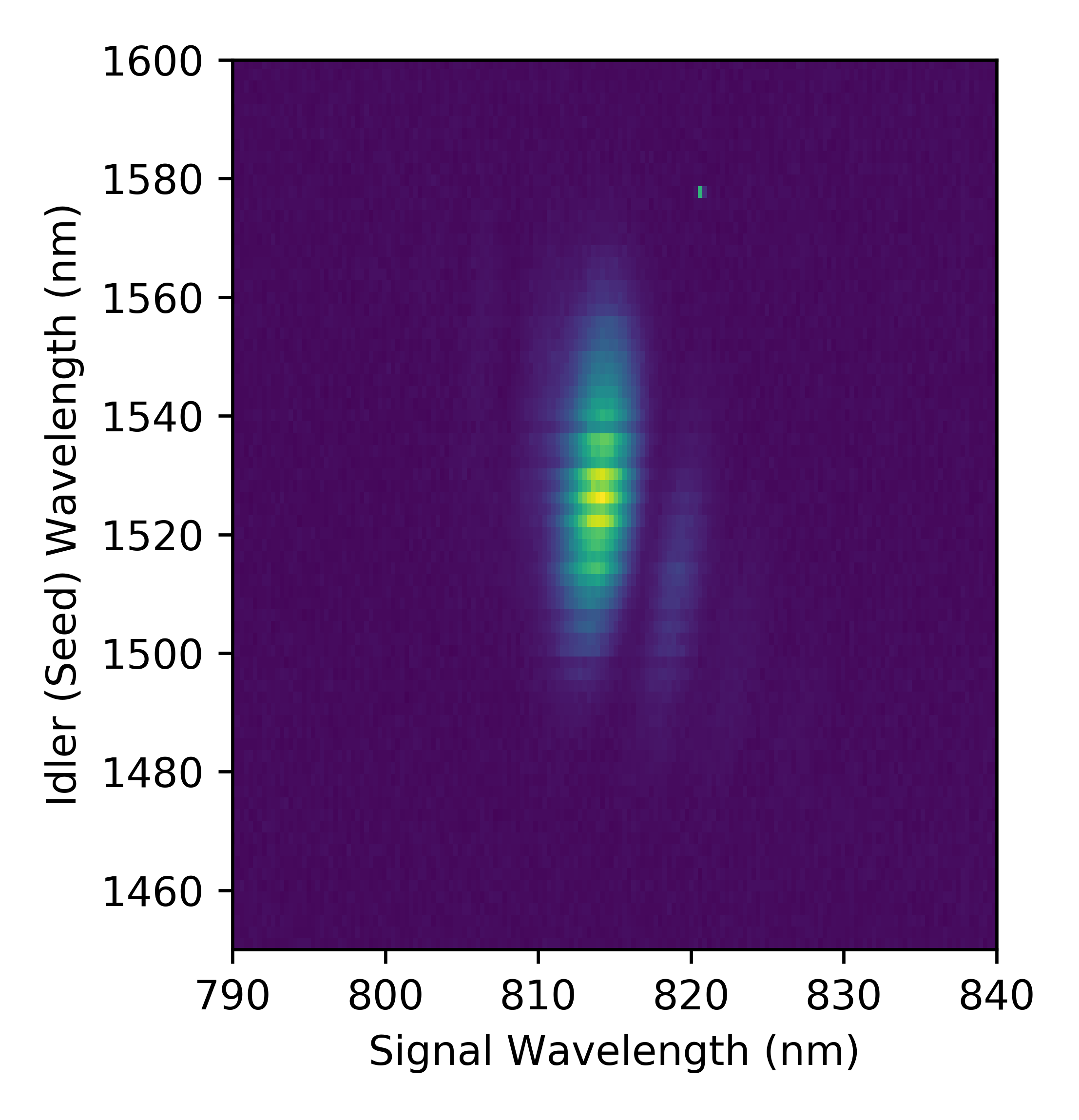}
  \end{subfigure}
\begin{subfigure}[r]{0.5\textwidth}
\caption{}\label{fig:JSI2}
  \includegraphics[width=0.85\textwidth]{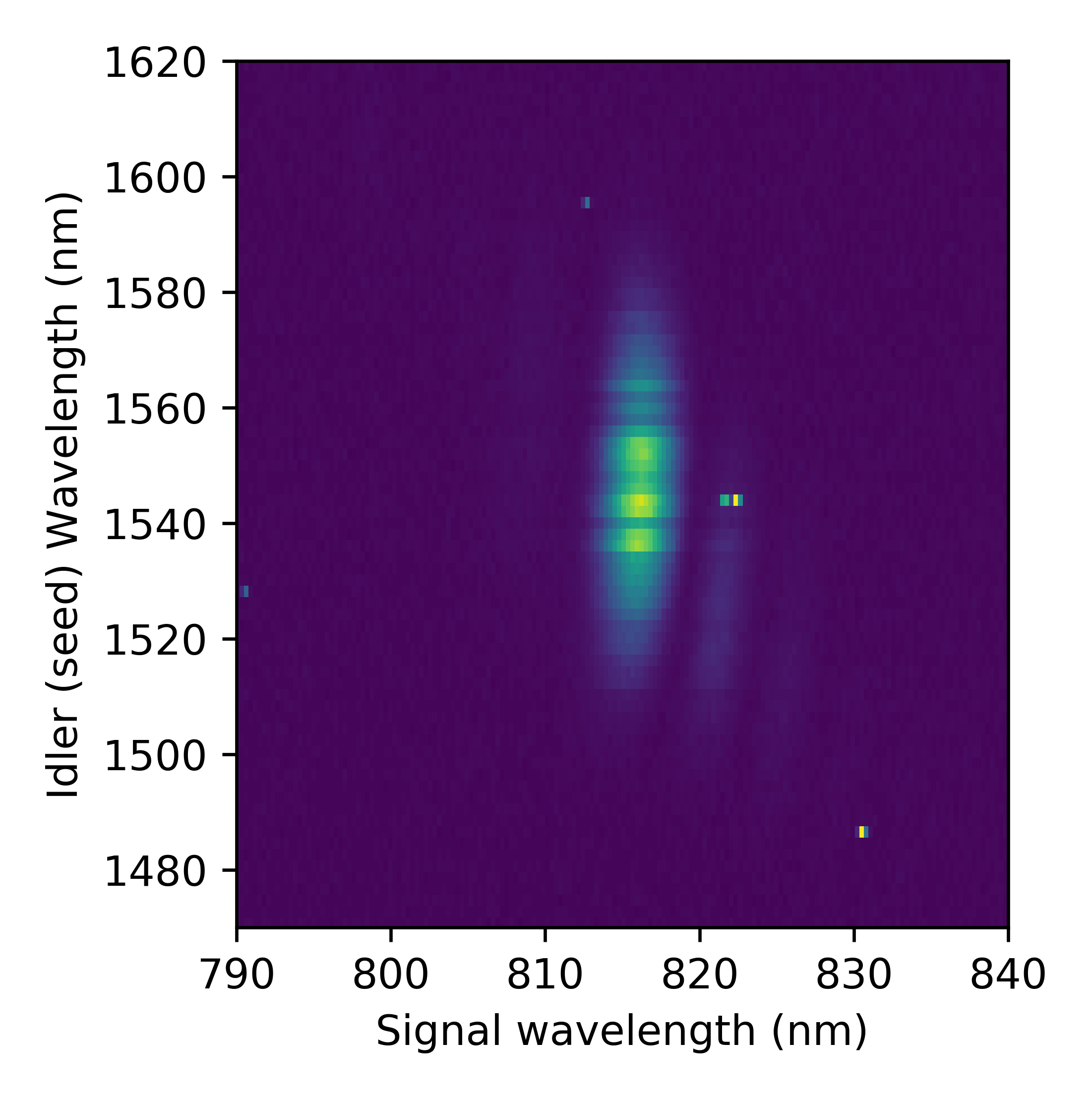}
\end{subfigure}
\caption{Seeded joint spectral intensity for two chips from wafer v2. (\subref{fig:JSI1}) Chip 1 ($\Delta n = \num{4.5+-0.2e-4}$) and (\subref{fig:JSI2}) chip 2 ($\Delta n = \num{4.9+-0.2e-4}$). 
Apparent horizontal bands are measurement artifacts from the
stacking of narrow-band seeded sections of the JSI.
}
\label{fig:JSI}
\end{figure*}

The experimental setup for the seeded JSI measurement is shown in Fig. \ref{fig:setups} (b). The pump source was a pulsed tunable laser 
operating at a repetition rate of \SI{250}{kHz} with a filtered bandwidth of \SI{20}{nm} centered between 1060 and \SI{1065}{nm}; the preparation was adapted from a similar source setup discussed in Ref. \cite{Spring2013}. A continuous-wave tunable telecommunications laser was used as the seed signal.   
Lenses were used for coupling signals on and off chip, along with a dichroic mirror and an edge-pass filter for the separation of pump and seed fields and further pump suppression, respectively.
The signal was coupled to single-mode fiber and analyzed on a spectrometer.
Figures \ref{fig:JSI} (\subref{fig:JSI1}) and (\subref{fig:JSI2}) show the seeded JSI measurements for two DUW waveguides in chips made from wafer v2. The similar spectral properties indicate that the SFWM process is possible with chips from different areas of the same wafer; local chip birefringence can be compensated by tuning the pump wavelength to obtain signals in the desired spectral regions.  
The vertical orientation of the JSI ellipse shows that the signal and idler frequency modes are approximately separable\cite{Grice2001}, thus a pure heralded state with little spectral filtering can be achieved by birefringent phase-matched SFWM in high-birefringence waveguides. In addition, this spectral-separability permits  filtering of the side lobes in the heralding arm to further increase the heralded single-photon purity without significant deterioration of the source brightness~\cite{Spring2017}.

The fabricated chips have been designed for waveguides capable of supporting single mode operation at the wavelengths of the pump (\SI{1060}{nm}), signal (\SI{790}{nm}) and idler (\SI{1550}{nm}), over which range there are expected variations of the coupling efficiency to a given optical fiber. To assess the fiber-to-waveguide coupling efficiency, an optical measurement of the $1/e^2$ Gaussian fit of the mode-field diameter (MFD) of the waveguide was done using a highly sensitive linear camera 
and a superluminescent light emitting diode 
The waveguide MFD for the fabricated chips was measured and was \SI{3.6}{\um} horizontally and \SI{3.7}{\um} vertically at \SI{790}{\nm}. The measured parameters were used to develop a waveguide model in a numerical mode solver package 
(Photon Design FIMMWAVE). The same waveguide geometry was used to simulate the waveguide mode field profile at \SI{1060}{nm} and \SI{1550}{nm}, as no direct measurements were available at these wavelengths. A geometry for Nufern PM-980-HP has also been implemented in 
FIMMWAVE based on the manufacturer's specification sheet (MFD \SI{5.5 +- 0.5}{\um} at \SI{980}{nm}). 
The theoretical fiber-to-waveguide coupling efficiency was \SI{87}{\percent} at \SI{1060}{nm}, \SI{91}{\percent} at \SI{790}{nm} and \SI{78}{\%} at \SI{1550}{nm}, which is in good agreement with experimental coupling measurements in the C-band.
The devices therefore provide efficient fiber-coupling of all fields on and off of the chip, permitting integration of single-photon source arrays into quantum optics experiments. 

\section{Discussion}

The proposed fabrication process demonstrates the suitability of utilizing the intrinsic high-birefringence of FHD planar layers with DUW for channel waveguide fabrication to  
generate photon pairs that are uncorrelated in frequency, with separation of the signal and idler from the pump of roughly \SI{300 }{nm}. The large spectral range of signal, idler and pump wavelengths in this scheme was found to induce wavelength-dependent coupling efficiency. 
For applications that require increased coupling efficiency of the idler field, this could be readily achieved by the fabrication of thicker core layers through the FHD deposition process. Thicker FHD core layers will support higher-order modes at the lower wavelength range; however, phase matching conditions for higher order modes are expected to be spectrally offset, permitting spectral filtering without degradation of the source brightness.

Chip-to-chip variation of the birefringence has been observed, though this could be compensated by using slightly different pump wavelengths for each chip. Additional processing, such as etching, may allow apodization of the waveguide birefringence profile to provide an optimized spectral output. The platform with chips of different birefringence permits integration with other DUW-based functionality, such as pump splitters and Bragg grating filters and cavities, for the development of other photonic integrated circuits elements. 

\section{Conclusion}

This paper reports on the suitability of the flame hydrolysis deposition process in conjunction with direct UV writing to fabricate waveguides with birefringence up to \num{4.9+-0.2e-4}. Grating-based characterization has been used to measure the birefringence of on-chip integrated channel waveguides. Seeded joint-spectral intensity measurements indicate low spectral correlations in phase-matched SFWM, with the ability for spectral filtering without significant deterioration of source brightness. Fiber-to-waveguide coupling efficiencies of \SI{87}{\%} have been modeled between the fabricated waveguides and commercial PM-fiber near pump wavelengths of \SI{1060}{\nm}, offering a route towards further stability and packaging of these devices in fiber networks. This investigation offers promising future perspectives for potential applications in photonic quantum experiments. 

\section*{Funding}

Engineering and Physical Sciences Research Council (EPSRC), UK grants EP/M013243/1, EP/M013294/1, EP/M024539/1, EP/K034480/1, \& 1375564.

\section*{Acknowledgements}

Mr.~A.~Jantzen for Fig.~\ref{fig:fab} (b). All data supporting this study are openly available from the University of Southampton repository at \url{https://doi.org/10.5258/SOTON/D0511}.

\bibliography{biblio} 
\bibliographystyle{osajnl}
\end{document}